\documentclass[aps,twocolumn,prl]{revtex4-2}

\usepackage{amsmath,graphicx,color}
\usepackage{bm}
\usepackage{physics}

\usepackage{bbold}
\usepackage{lipsum}
\usepackage{natbib}
\usepackage{placeins}
\usepackage{braket}

\begin{document}
\title{Measuring the Debye Energy in Superconductors via two Electron Photoemission Spectroscopy}
\author{Ka Ho Wong$^{1}$, Jack Zwettler$^{2,3}$, Henry Amir$^{2,3}$, Peter Abbamonte$^{2,3}$, Fahad Mahmood$^{2,3}$, and Dirk K. Morr$^{1}$}
\affiliation{$^{1}$ Department of Physics, University of Illinois at Chicago, Chicago, IL 60607, USA}
\affiliation{$^{2}$ Department of Physics, University of Illinois at Urbana-Champaign, Urbana, IL 61801, USA}
\affiliation{$^{3}$ Materials Research Laboratory, University of Illinois at Urbana-Champaign, Urbana, IL 61801, USA}

\begin{abstract}
We demonstrate theoretically that double angle resolved photoemission spectroscopy (2eARPES) can directly probe the existence of Cooper pairs away from the Fermi surface, and can thus provide insight into the characteristic energy scale around the Fermi surface, the Debye energy, in which electrons are bound into Cooper pairs. To this end, we compute the photoelectron counting rate $P^{(2)}$ in two different types of unconventional superconductors, a $d_{x^2-y^2}$-wave superconductor, and a topological superconductor with a broken time-reversal symmetry.  We show that $P^{(2)}$ provides insight into the relative strength of intra- and inter-band pairing in multi-band systems, as well as into the spin polarization of the bands.
\end{abstract}

\maketitle

{\it Introduction}
The Debye energy, $\omega_D$, is a central concept in the study of superconductors \cite{BCS1957,Tinkham2004} as it represents the energy range around the Fermi energy in which electrons are bound into Cooper pairs. However, gaining experimental insight into this elusive quantity, and in general, into the existence of Cooper pairs away from the Fermi surface, has proven extremely difficult.
For example, while angle-resolved photo-emission experiments can in general map out the momentum dependence of the superconducting order parameter $|\Delta_{\bf k}|$ by measuring the induced shift in the electronic normal state dispersion \cite{Sobota2021}, the experimental resolution often confines this approach to the immediate vicinity of the Fermi surface. In contrast, it was shown that the photo-electron counting rate, $P^{(2)}$, in double-ARPES (2e-ARPES) experiments, in which one photon leads to the emission of 2 electrons, directly reflects the existence of Cooper pairs \cite{Berakdar2003,Berakdar2006,Mahmood2022,wong2eARPES,KOUZAKOV2007121,DPE_theory}, and provides insight into their spin and momentum structure. The question thus naturally arises of whether it can also probe the energy and momentum range away from the Fermi surface in which superconducting pairing occurs.

\begin{figure}[htb]
\center
\includegraphics[width=0.95\linewidth]{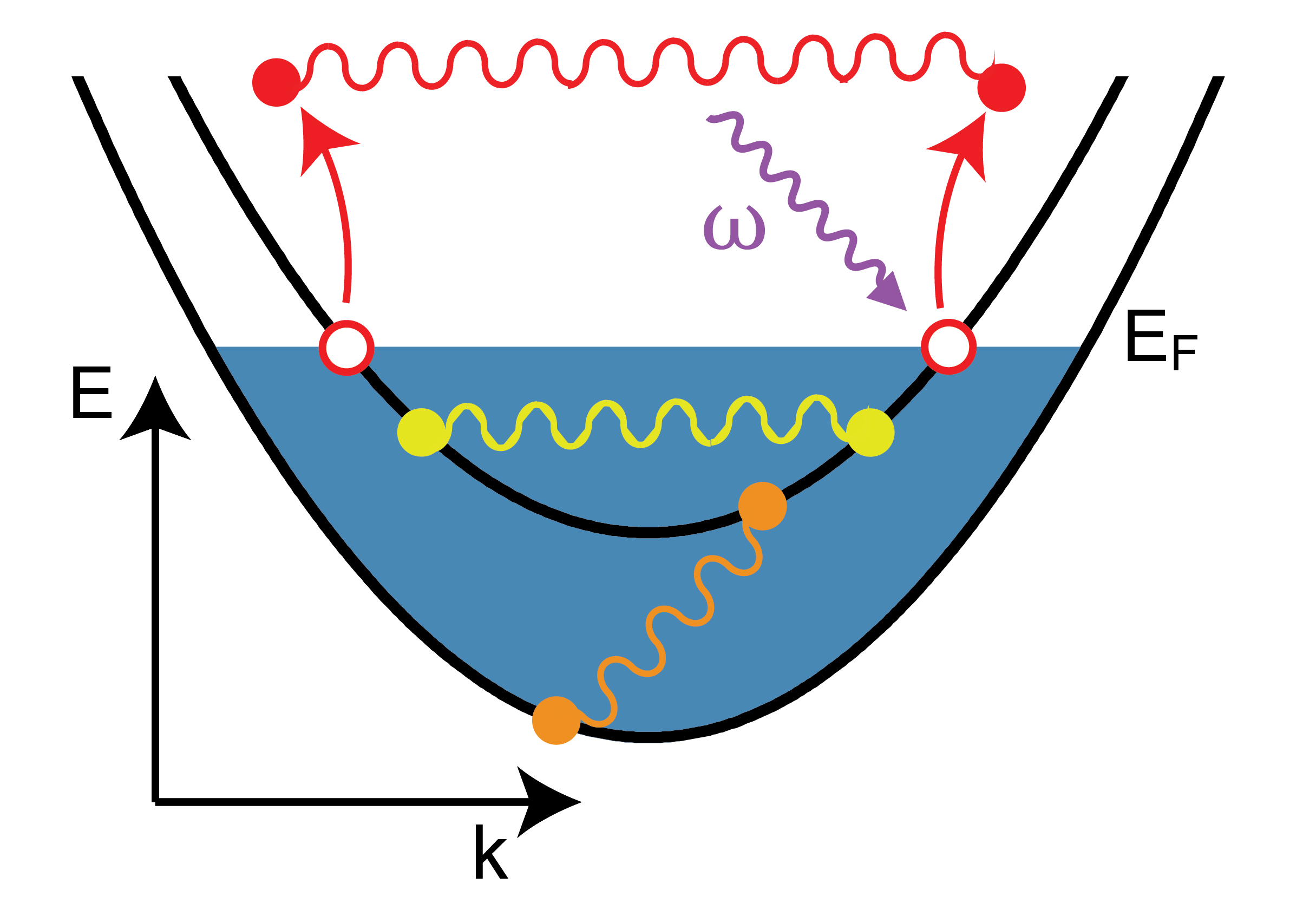}
\caption{Schematic representation of the emission of two entangled photo-electrons from a Cooper pair at the Fermi surface (red circles), and from Cooper pairs located away from the Fermi surface with yellow (orange) circles representing superconducting intra-band (inter-band)  pairing.}
\label{fig:Fig0}
\end{figure}

In this Letter, we theoretically demonstrate that 2e-ARPES \cite{Trutzschler2017,Chiang2020,Schumann2020,ZWETTLER2024147417} can directly probe the existence of Cooper pairs away from the Fermi surface, and thus provide insight into the characteristic energy range around the Fermi surface
-- the Debye energy -- in which Cooper pairs are formed. In particular, we show that the contribution to the photo-electron counting rate, $P^{(2)}$, of 2eARPES experiments that directly measures the emission of two entangled photo-electrons from the same Cooper pair, remains non-zero for all momenta where Cooper pairs exist. To demonstrate this, we study an unconventional, single-band  $d_{x^2-y^2}$-wave superconductor, as exemplified by the cuprate superconductors \cite{Cuprate_review,Vishik_2018}, as well as a multi-band topological superconductor with a broken time-reversal symmetry \cite{Zhang2018,MgB2_topological_SC,2M-WS2_topological_SC,SR2RuO4_topological_SC}. We demonstrate that $P_{SC}^{(2)}$ for both systems does not only directly probe the Debye energy, but can also provide insight into the relative strength between superconducting inter- and intra-band pairing, as well as the spin-polarization of the involved electronic bands, as schematically shown in Fig.~\ref{fig:Fig0}. Our results thus demonstrate that 2e-ARPES experiments open an unprecedented opportunity to gain insight into one of the most fundamental aspects of superconductivity.

{\it Theoretical Model}
In the following, we demonstrate that the photo-electron counting rate $P^{(2)}$ measured in 2$e$-ARPES provides direct insight into the elusive Debye energy for two different types of superconductors: (i) an unconventional $d_{x^2-y^2}$-wave superconductor, representative of the cuprate superconductors, and (ii) a multi-band topological superconductor with broken time reversal symmetry -- the 2DTSC model --which was proposed as an explanation for the topological phase of FeSe$_{x}$Te$_{1-x}$ \cite{Zhang2018,Mascot2022,Xu2023}. The former is described by the Hamiltonian
\begin{align}
H_d&=\sum_{\vb{k}, \sigma}\xi_{\vb{k}}c^{\dagger}_{\vb{k},\sigma}c_{\vb{k},\sigma}
-\sum_{\abs{\xi_{\vb{k}}}<\omega_{D}}\Delta_{\vb{k}} \qty(c^{\dagger}_{\vb{k},\uparrow}c^{\dagger}_{-\vb{k},\downarrow}+H.c.)
\end{align}
Here, $c^{\dagger}_{\vb{k},\sigma}$ creates an electron with momentum ${\bf k}$ and spin $\sigma$, $\xi_{\vb{k}}$
is the normal state tight binding dispersion,
$\Delta_{\vb{k}}=\frac{\Delta_0}{2}\qty(\cos{k_x}-\cos{k_y})$ is the $d_{x^2-y^2}$-wave superconducting order parameter and $\omega_{D}$ is the Debye energy [for details, see Supplementary Material (SM) Sec.~I].\\

The Hamiltonian of the 2DTSC model is given by \cite{Rachel2017,Mascot2022,Xu2023}
\begin{align}
H_{SC}&=\sum_{\vb{k},\sigma}\xi_{\bf k}
c^{\dagger}_{\vb{k},\sigma}c_{\vb{k},\sigma} - \sum_{\vb{k}} \Delta_{0}\qty(c^{\dagger}_{\vb{k},\uparrow}c^{\dagger}_{-\vb{k},\downarrow}+H.c.)\nonumber \\
&\qquad +2\alpha\sum_{\vb{k},\sigma,\sigma'}\qty(\sin{k_x}\sigma^{y}_{\sigma\sigma'}-\sin{k_y}\sigma^{x} _{\sigma\sigma'})c^{\dagger}_{\vb{k},\sigma}c_{\vb{k},\sigma'} \nonumber \\
&\qquad-JS\sum_{\vb{k},\sigma,\sigma'}c^{\dagger}_{\vb{k},\sigma}\sigma^{z}_{\sigma\sigma'}c_{\vb{k},\sigma'}  \ ,
\end{align}
where $\xi_{\bf k}$ is the tight-binding dispersion, $\Delta_0$ is the $s$-wave superconducting order parameter, $\alpha$ is the Rashba spin-orbit (RSO) interaction, and $J$ is the magnetic exchange coupling between the ordered moments of magnitude $S$ and the conduction electrons (the implementation of the Debye energy for this model is discussed in SM Sec.~II).

The emission of two correlated photoelectrons in 2$e$ARPES experiments is a two-step process: the absorption of a photon leads to the emission of a first photo-electron that subsequently interacts via the Coulomb interaction with a conduction electron, leading to the emission of a second photoelectron \cite{Haak1978,Berakdar1998,Chiang2020,Fominykh2002,Schumann2007,Schumann2011,Schumann2012,Herrmann1998,Mahmood2022,wong2eARPES}. These two processes are described by the Hamiltonian \cite{Mahmood2022,wong2eARPES}
\begin{align}
    H_{scat} & = \sum_{{\bf k,q},\sigma,\nu} \gamma_{\nu}({\bf q}) d^\dagger_{{\bf k+q},\sigma} c_{{\bf k},\sigma} \left( a_{{\bf q},\nu}  + a^\dagger_{-{\bf q},\nu}\right)  \nonumber \\
    & + \sum_{{\bf k,p,q},\alpha, \beta} V({\bf q}) d^\dagger_{{\bf k}+{\bf q},\alpha} d^\dagger_{{\bf p}-{\bf q},\beta} d_{{\bf p},\beta} c_{{\bf k},\alpha}  + H.c.
    \label{eq:Ham}
\end{align}
Here, $\gamma_{\nu}({\bf q})$ is the effective electron-photon dipole interaction, $d^\dagger_{{\bf k},\sigma} (c_{{\bf k},\sigma})$ creates (destroys) a photo-electron (conduction electron) with momentum ${\bf k}$ and spin $\sigma$, and $V({\bf q}) = V_0/ \left({\bf q}^2 + \kappa^{2}\right)$ is the Fourier transform of the (screened) Coulomb interaction, with $\kappa^{-1}$ being the screening length. Since the photon momentum is much smaller than typical fermionic momenta, we set it equal to zero, such that $\gamma_{\nu}({\bf q}) = \gamma_0$ is simply a momentum-independent constant. As previously shown \cite{Mahmood2022,wong2eARPES}, the photo-electron counting rate resulting from this interaction can be written as $P^{(2)}=V^2 P^{(2)}_{SC}+V P^{(2)}_{2cp}$ with $V$ being the volume of the system. The first term arises from the emission of two correlated photoelectrons from a single Cooper pair, and thus directly reflects the existence of a superconducting condensate, while the second term describes the emission of two photoelectrons from two different Cooper pairs. We note that $P^{(2)}$ is a function of both the momenta, ${\bf k}_{1,2}^\prime$ and the spins, $\sigma_{1,2}$ of the two photo-electrons. \\

{\it Results}
\begin{figure}[htb]
\center
\includegraphics[width=\linewidth]{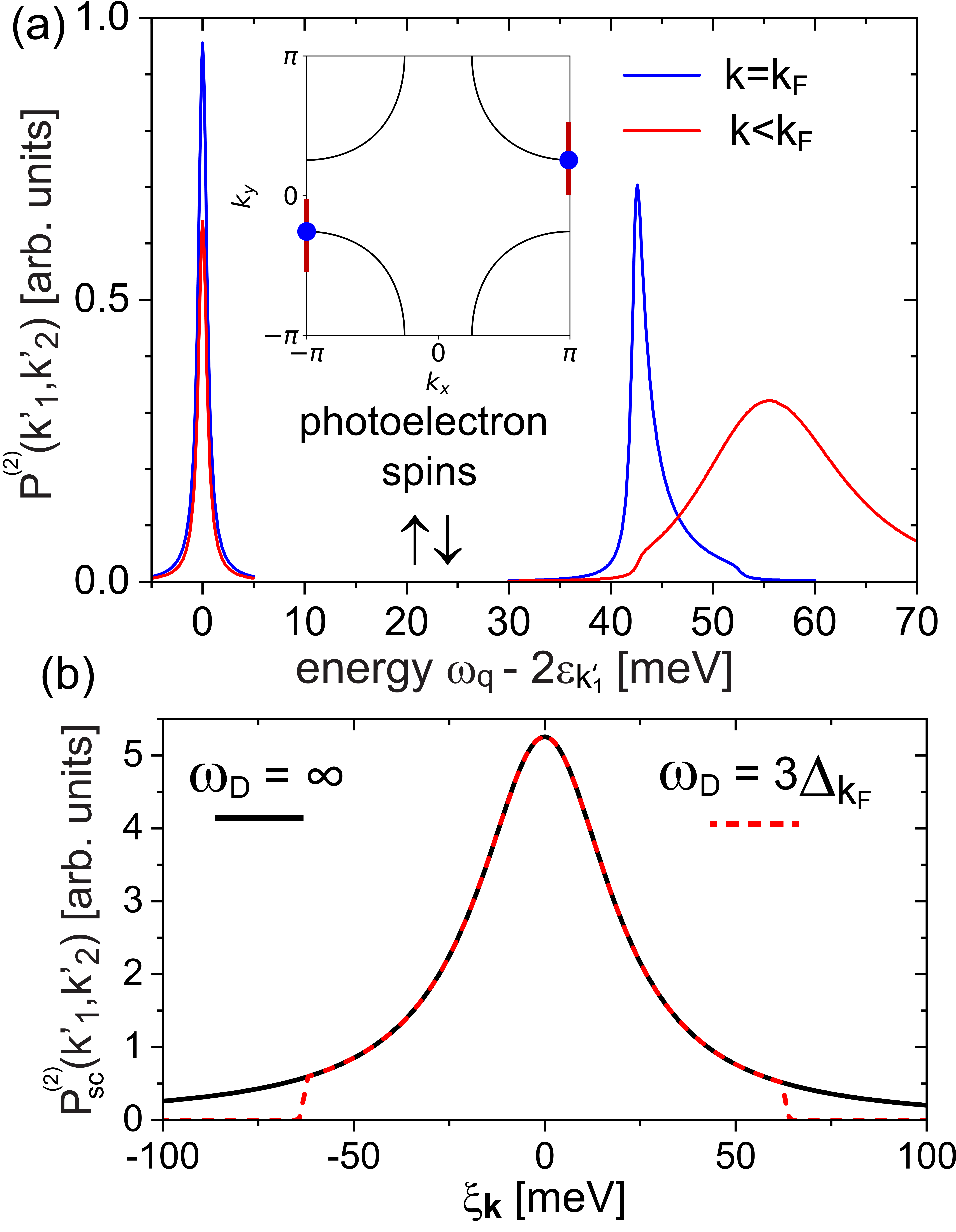}
\caption{(a) $P^{(2)}$ in a $d_{x^2-y^2}$-wave superconductor for two photo-electrons with opposite spin and opposite momenta ${\bf k}^\prime_2=-{\bf k}^\prime_1$ at (see filled blue circles in the inset) or below the Fermi surface. Inset: Fermi surface in the normal state. (b) $P^{(2)}_{SC}$ as a function of the normal state energy, $\xi_{\bf k}$, along a line cut in the Brillouin zone indicated by a red line in the inset of (a). Black line: $\omega_D = \infty$, red line: $\omega_D = 3 \Delta_{\vb{k_F}}$. Parameters used are $(t,t',\mu,\Delta_0)=(300,-120,-150,25)\mathrm{meV}$.}
\label{fig:Fig1}
\end{figure}
We begin by considering the photo-electron counting rate $P^{(2)}$ in a $d_{x^2-y^2}$-wave superconductor, for which the Fermi surface in the normal state is shown in Fig.~\ref{fig:Fig1}(a). For two photo-electrons with opposite momenta ${\bf k}^\prime_2=-{\bf k}^\prime_1$ lying on the Fermi surface, as indicated by filled blue circles in the Fig.~\ref{fig:Fig1}(a), and opposite spins, the resulting photo-electron counting rate $P^{(2)}$ exhibits a peak at $\Delta \omega = \omega_{q}-2\epsilon_{\vb{k}_1'}=0$ arising from $P^{(2)}_{SC}$, as shown in Fig.~\ref{fig:Fig1}(b). Here, $\omega_{q}$ is the photon energy, and $\epsilon_{\vb{k}_1'}$ is the sum of the kinetic energy of a photo-electron and the work-function $W$; thus $\Delta \omega$ represents the excess energy of the photon over the energies of the two photo-electrons. In addition, $P^{(2)}$ exhibits a continuum with onset at  $\Delta \omega \approx 2\Delta_{{\bf k}_1^\prime}$ arising from $P^{(2)}_{2cp}$.

It is generally assumed that electrons form Cooper pairs not only at or close to the Fermi surface, but over an extended energy range around the Fermi energy, $E_F$, generally known as the Debye energy. We thus expect that the zero-energy peak persists even for photo-electron momenta away from the Fermi surface, until the normal state energy of the emitted conduction electrons crosses the Debye energy. This is confirmed by the calculation of $P^{(2)}$ for two photo-electrons with momenta away from the Fermi surface for which the zero-energy peak persists, though with reduced intensity, while the onset of the continuum shifts to higher energies, as shown in Fig.~\ref{fig:Fig1}(a). To further demonstrate this, we plot in  Fig.~\ref{fig:Fig1}(b) $P^{(2)}_{SC}$ for a linecut along the boundary of the Brillouin zone (BZ) [see red lines in the inset of Fig.~\ref{fig:Fig1}(a)] that crosses the Fermi surface.
We find that $P^{(2)}_{SC}$ exhibits a maximum for momenta at the Fermi surface, i.e.  ${\bf k}_1^\prime = {\bf k}_F$ and thus $\xi_{{\bf k}_1^\prime} = 0$, and decreases with increasing distance from the Fermi surface. This momentum dependence can be approximately described by (see SM Sec.~III)
\begin{equation}
    \frac{P^{(2)}_{SC}(\vb{k}_1')}{P^{(2)}_{SC}(\vb{k}_{F})} \approx \frac{{\Delta_{\vb{k}_1'}^2}}{{\Delta_{\vb{k}_1'}^2}+\xi_{\vb{k}_1'}^2} \ .
    \label{eq:P2SC}
\end{equation}
This ability of 2eARPES experiments to probe the existence of Cooper pairs away from the Fermi surface can be employed to identify the Debye energy. To demonstrate this, we compare in Fig.\ref{fig:Fig1}(b)  $P^{(2)}_{SC}$ for the case of  $\omega_D=\infty$ and $\omega_D=3\Delta_{{\bf k}_F}$, where $\Delta_{{\bf k}_F}$ is the superconducting gap at the Fermi surface. We find that for the latter case, $P^{(2)}_{SC}$ drops to zero for $|\xi_{\bf k}|>\omega_D$ as no Cooper pairs are formed for larger energies. This establishes the proof of concept that 2eARPES experiments can determine the energy scale beyond which Cooper pairs cease to exist.

\begin{figure}[htb]
\center
\includegraphics[width=\linewidth]{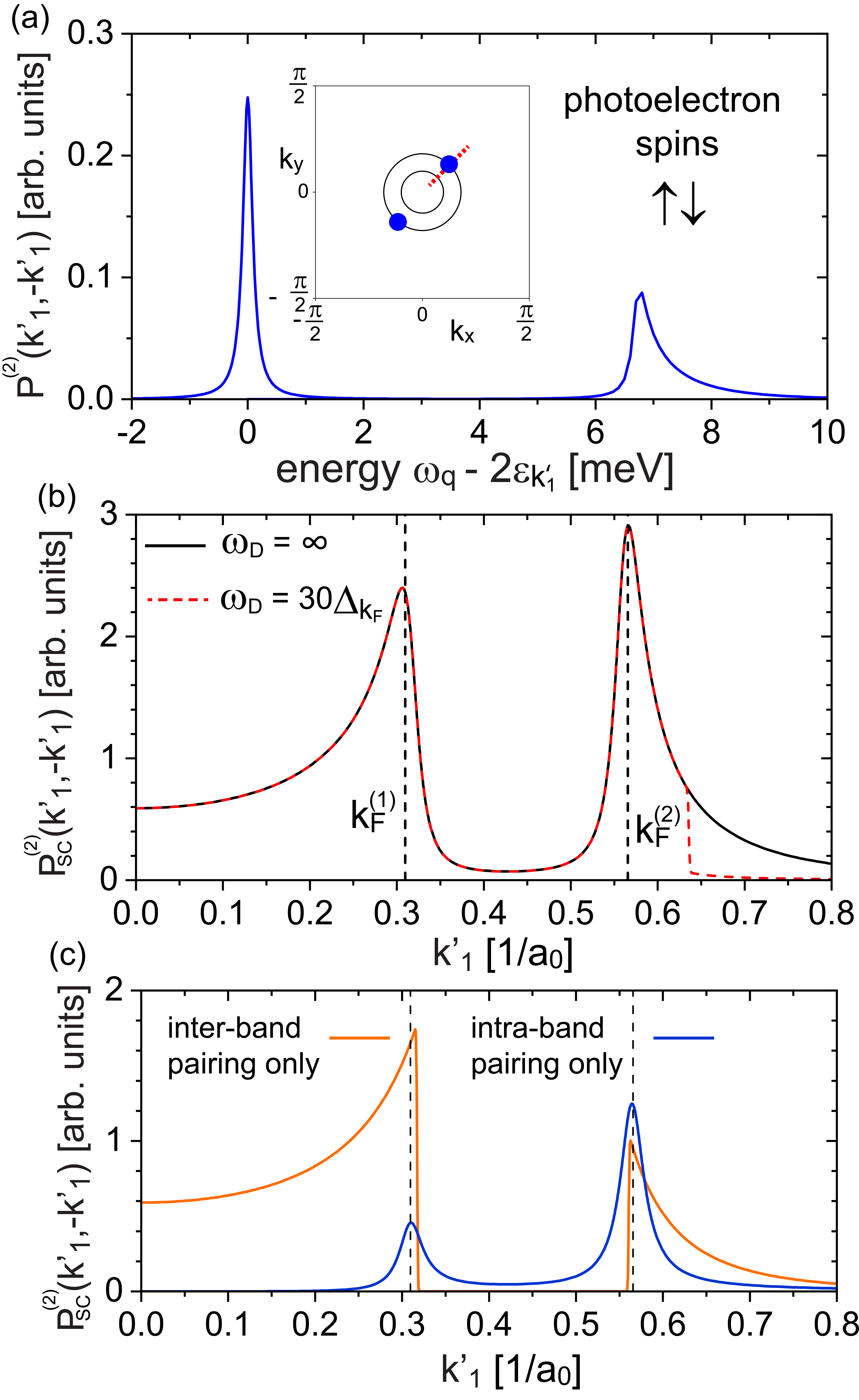}
\caption{(a) $P^{(2)}$ as a function of $\Delta \omega$ for two photoelectrons with opposite spin and momentum and located at the outer Fermi surface (see filled blue circles in the inset). Inset: normal state Fermi surfaces. (b) $P^{(2)}_{SC}$ as a function of ${\bf k}_1^\prime$ along the diagonal of the BZ for two values of $\omega_D$. (c) $P^{(2)}_{SC}$ assuming intra-band (blue line) and inter-band (orange line) pairing only. Parameters used are $(t,\mu,\alpha,\Delta_0,JS)=(200,-760,10,7,20)\mathrm{meV}$.}
\label{fig:Fig2}
\end{figure}
We next study the form of the photo-electron counting rate in the 2DTSC system, which is of particular interest for two reasons. First, it is a multi-band system, thus allowing not only for superconducting intra-band, but also for inter-band pairing. Second, the interplay between the RSO interaction and the presence of ferromagnetism leads to the emergence of superconducting spin-triplet correlations, as well as spin-polarized bands \cite{Rachel2017}. The question thus naturally arises of whether both of these features lead to characteristic signatures in $P^{(2)}_{SC}$. To address this question, we begin by considering a parameter set for which the system is in the topologically trivial phase, and possesses two nearly isotropic Fermi surface closed around the $\Gamma$ point, as shown in the inset of Fig.~\ref{fig:Fig2}(a). The resulting photo-electron counting rates for two photo-electrons of opposite momenta ${\bf k}_2^\prime = - {\bf k}_1^\prime$ on the outer Fermi surface [see filled blue circles in the inset of Fig.~\ref{fig:Fig2}(a)] and opposite spins is shown in Fig.~\ref{fig:Fig2}(a). As for the case of the $d_{x^2-y^2}$-wave superconductor, $P^{(2)}$ exhibits a peak at $\Delta \omega=0$ arising from $P^{(2)}_{SC}$ which is separated from the onset of the continuum, arising from $P^{(2)}_{2cp}$, by $2\Delta_{{\bf k}_1^\prime}$. In Fig.~\ref{fig:Fig2}(b), we present $P^{(2)}_{SC}$ for a linecut along the diagonal direction in the BZ [see dashed red line in the inset of Fig.~\ref{fig:Fig2}(a)] that crosses both Fermi surfaces at $k^{(1,2)}_F$. In contrast to the case for a $d_{x^2-y^2}$-wave superconductor [cf.~Fig.~\ref{fig:Fig1}(b)], $P^{(2)}_{SC}$ exhibits a strongly anisotropic shape around the Fermi momenta $k^{(1,2)}_F$ (denoted by vertical black dashed lines). To identify the origin of this asymmetry, we consider in Fig.~\ref{fig:Fig2}(c) the form of $P^{(2)}_{SC}$ for two limiting cases: when inter-band pairing is suppressed and superconductivity arises from intra-band pairing only [see blue line in Fig.~\ref{fig:Fig2}(c)], and vice versa (orange line). In the former case, $P^{(2)}_{SC}$ is nearly symmetric around  $k^{(1,2)}_F$, and hence similar to the case of the single-band $d_{x^2-y^2}$-wave superconductor. In contrast, for intra-band pairing only, $P^{(2)}_{SC}$ is highly asymmetric around  $k^{(1,2)}_F$, exhibiting a sharp momentum cut-off, and a finite range of momenta where it vanishes, thus indicating the absence of superconducting correlations. To understand this form of $P^{(2)}_{SC}$, we recall that inter-band pairing can only  occur when the time-reversed states that are being paired, $E^{(1)}_{\bf k}$ and $E^{(2)}_{-{\bf k}}$, are either both occupied or both unoccupied. The sharp cut-off in $P^{(2)}_{SC}$ then occurs at those momenta, where one of the bands crosses the Fermi energy, as it is not possible to sustain superconducting correlations when the time-reversed and hole-like partner of a particle state at ${\bf k}$ does not exist (for a more detailed discussion, see SM Sec.~IV). A comparison of the results in Figs.~\ref{fig:Fig2}(b) and \ref{fig:Fig2}(c) then shows that the asymmetry in $P^{(2)}_{SC}$ shown in Fig.~\ref{fig:Fig2}(b) arises from the inter-band pairing term, which thus possesses a characteristic signature in $P^{(2)}_{SC}$ that can be used to identify it. We note that $P^{(2)}_{SC}$ for inter-band pairing decreases much more slowly with distance from the Fermi momenta than that for intra-band pairing due to the different form of the coherence factors involved (for details, see SM Sec.~IV). Thus, $P^{(2)}_{SC}$ exhibits considerable intensity even for momentum states with energies much larger than the superconducting gap, thus facilitating the detection of much larger Debye energies, as shown for the case of $\omega_D = 30 \Delta_{{\bf k}_F}$ in Fig.~\ref{fig:Fig2}(b).

\begin{figure}[htb]
\center
\includegraphics[width=\linewidth]{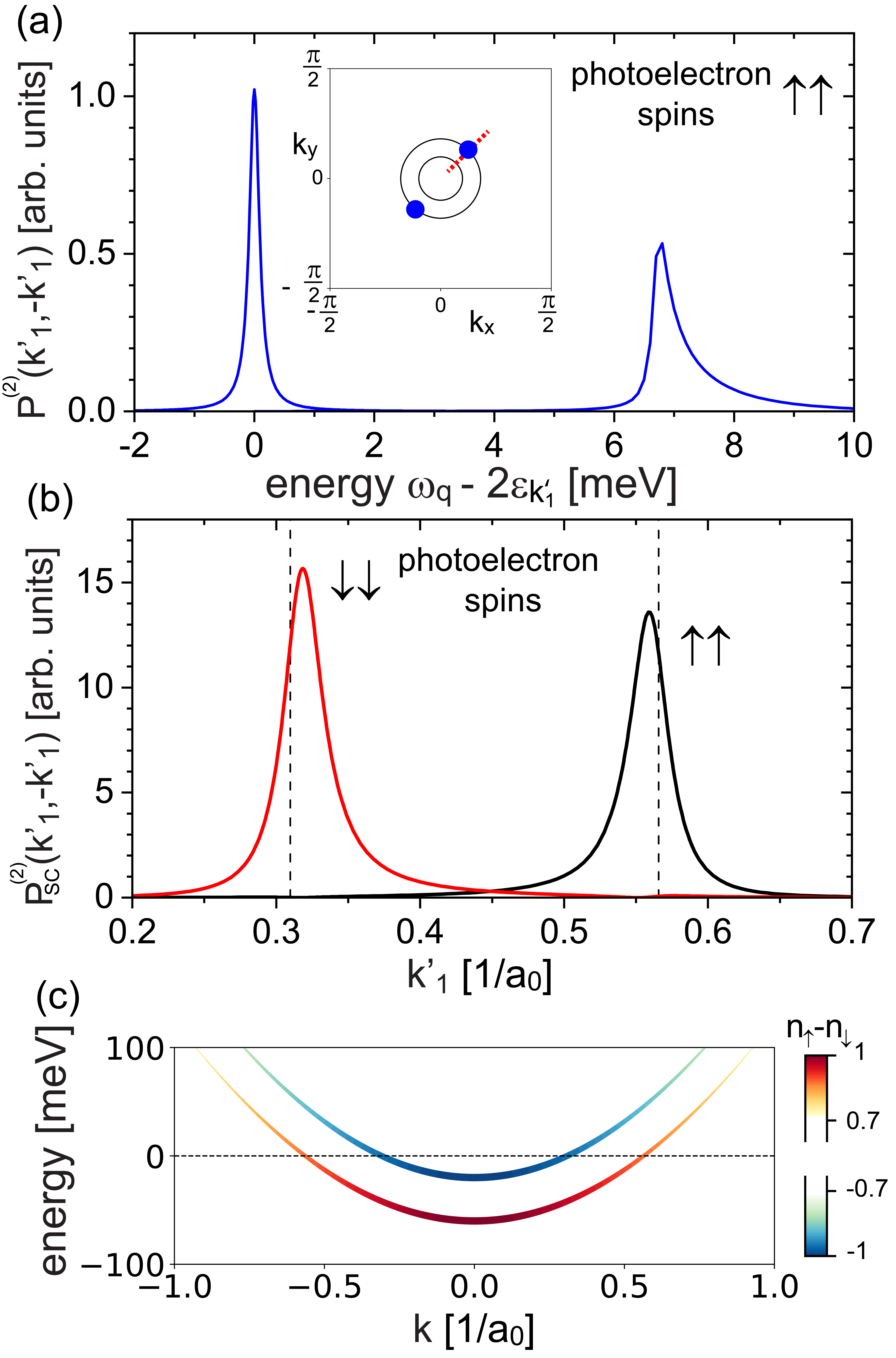}
\caption{(a) $P^{(2)}$ as a function of $\Delta \omega$ for two photoelectrons with the same spins and opposite momenta, located at the outer Fermi surface (see filled blue circles in the inset). Inset: normal state Fermi surfaces. (b) $P^{(2)}_{SC}$ as a function of ${\bf k}_1^\prime$ along the diagonal of the BZ for photoelectrons with $\uparrow \uparrow$ and $\downarrow \downarrow$ spin configurations. (c) Spin polarization of the electronic bands in the normal state. Parameters used are $(t,\mu,\alpha,\Delta_0,JS)=(200,-760,10,7,20)\mathrm{meV}$.}
\label{fig:Fig3}
\end{figure}

We next consider the form of $P^{(2)}$ for opposite momenta, but equal spin of the photo-electrons, thus probing the existence of superconducting spin-triplet correlations. In Fig.~\ref{fig:Fig3}, we present $P^{(2)}$ as a function of $\Delta \omega$ for two momenta on the Fermi surface [see filled blue circles in the inset of Fig.~\ref{fig:Fig3}(a)].
As previously discussed \cite{Rachel2017}, the combination of ferromagnetism, RSO interaction and an $s$-wave gap leads to the emergence of superconducting spin-triplet correlations, which are reflected in the presence of a peak at $\Delta \omega = 0$ arising from $P^{(2)}_{SC}$. Plotting $P^{(2)}_{SC}$ for the $\uparrow \uparrow$ ($S_z=+1$) and $\downarrow \downarrow$ ($S_z=-1$) spin configurations of the two photoelectrons for ${\bf k}_1^\prime$ along the BZ diagonal [see dashed red line in the inset of Fig.~\ref{fig:Fig3}(a)], we observe two interesting features. First, $P^{(2)}_{SC}$ for either spin configuration is nearly symmetric around $k^{(1,2)}_F$, suggesting the absence of any appreciable inter-band pairing (see SM Sec.~V). Second, $P^{(2)}_{SC}$ exhibits an appreciable intensity in the $\downarrow \downarrow$-channel ($\uparrow \uparrow$-channel) only at ${\bf k}_F^{(1)}$ (${\bf k}_F^{(2)}$), i.e., at the inner (outer) Fermi surface. Both features directly arise from the strong and opposite spin-polarization of the two bands shown in Fig.~\ref{fig:Fig3}(c), which suppresses inter-band pairing, and allows for the emergence of superconducting spin-triplet correlations in the $(S_z=-1)$-channel only in the inner band, and in the ($S_z=+1$)-channel only in the outer band.  Finally, we note that similar results for $P^{(2)}_{SC}$ are also obtained for parameter sets where the system is in the topological phase, with the most significant effects arising from the much weaker spin polarization of the electronic bands (for a detailed discussion, see SM Sec.~VI). Thus, 2eARPES can not only probe the existence of intra- and inter-band pairing, but also of spin-polarized bands.

{\it Conclusions}
We have demonstrated that 2eARPES experiments can probe the existence of Cooper pairs away from the Fermi surface, thus allowing for the first time to explore the extent of the Debye energy. Moreover, we have shown that $P^{(2)}_{SC}$ exhibits a qualitatively different form for superconducting intra-band and inter-band pairing, providing an intriguing approach to identify and distinguish between these two pairing channels. Furthermore, the dependence of $P^{(2)}_{SC}$ in 2eARPES on the spin configuration of the photoelectrons also provides unique insight into the spin polarization of the electronic bands. Our results demonstrate that 2eARPES experiments thus can explore some of the most fundamental aspects of superconductivity. \\

{\bf Acknowledgements} This study was supported by the Center for Quantum Sensing and Quantum Materials, an Energy Frontier Research Center funded by the U. S. Department of Energy, Office of Science, Basic Energy Sciences under Award DE-SC0021238. J.Z, H.A., and F.M. acknowledge support from the EPiQS program of the Gordon and Betty Moore Foundation, Grant GBMF11069. P.A. acknowledges support from the EPiQS program of the Gordon and Betty Moore Foundation, Grant No. GBMF9452.

\end{document}

\title{Measuring the Debye Energy in Superconductors via two electron photoemission spectroscopy\\
Supplemental Information}
\author{Ka Ho Wong$^{1}$, Jack Zwettler$^{2,3}$, Henry Amir$^{2,3}$, Peter Abbamonte$^{2,3}$, Fahad Mahmood$^{2,3}$, and Dirk K. Morr$^{1}$}
\affiliation{$^{1}$ Department of Physics, University of Illinois at Chicago, Chicago, IL 60607, USA}
\affiliation{$^{2}$ Department of Physics, University of Illinois at Urbana-Champaign,Urbana,IL 61801, USA}
\affiliation{$^{3}$ Materials Research Laboratory, University of Illinois at Urbana-Champaign,Urbana,IL 61801, USA}

\maketitle

\section{Debye energy cut-off in a $d_{x^2-y^2}$-wave superconductor}

The Hamiltonian to describe a $d_{x^2-y^2}$-wave superconductor is given by
\begin{align}
H_d&=\sum_{\vb{k},\sigma}\xi_{\vb{k}}c^{\dagger}_{\vb{k},\sigma}c_{\vb{k},\sigma}
-\sum_{\vb{k}}\qty(\Delta_{\vb{k}} c^{\dagger}_{\vb{k},\uparrow}c^{\dagger}_{-\vb{k},\downarrow}+H.c.)
\label{eq:Hd}
\end{align}
Here, $c^{\dagger}_{\vb{k},\sigma}$ creates an electron with momentum ${\bf k}$ and spin $\sigma$, and $\Delta_{\vb{k}}=\frac{\Delta_0}{2}\qty(\cos{k_x}-\cos{k_y})$ is the $d_{x^2-y^2}$-wave superconducting order parameter. The normal state dispersion is given by
\begin{align}
    \xi_{\vb{k}} = -2t\qty(\cos{k_x}+\cos{k_y})-4t^\prime\cos{k_x}\cos{k_y}-\mu
\end{align}
with $-t (-t^\prime)$ being the nearest-neighbor (next-nearest-neighbour) hopping amplitude, $\mu$ is the chemical potential. For the results shown in the main text, we have used $t=300\mathrm{meV}$, $t^\prime = -120\mathrm{meV}$, $\mu=-150\mathrm{meV}$, and $\Delta_0 = 25\mathrm{meV}$.\\

To implement the Debye cut-off, we replace the superconducting order parameter in Eq.(\ref{eq:Hd}) by
\begin{align}
\Delta_{\vb{k}}\qty(\omega_{D})&=\Delta_{\vb{k}}\theta\qty(\omega_{D}-\abs{ \xi_{\vb{k}}})
\end{align}
where $\theta(x)$ is the Heaviside step function.\\

\section{Debye cut-off in the 2DTSC model}

The Hamiltonian  of the 2DTSC model is given by \cite{Rachel2017,Mascot2022}
\begin{align}
H_{SC}&=\sum_{\vb{k},\sigma}\xi_{\bf k}c^{\dagger}_{\vb{k},\sigma}c_{\vb{k},\sigma}
-JS\sum_{\vb{k},\sigma,\sigma^\prime}c^{\dagger}_{\vb{k},\sigma}\sigma^{z}_{\sigma\sigma^\prime}c_{\vb{k},\sigma^\prime}
   +2\alpha\sum_{\vb{k},\sigma,\sigma^\prime}\qty(\sin{k_x}\sigma^{y}_{\sigma\sigma^\prime}-\sin{k_y}\sigma^{x}_{\sigma\sigma^\prime})
c^{\dagger}_{\vb{k}, \sigma}c_{\vb{k},\sigma^\prime} -\sum_{\vb{k}}\Delta_{0}\qty(c^{\dagger}_{\vb{k},\uparrow}c^{\dagger}_{-\vb{k},\downarrow}+H.c.)\, .
\label{eq:HSC}
\end{align}
where
\begin{align}
 \xi_{\bf k} =    -2t\qty(\cos{k_x}+\cos{k_y})-\mu
\end{align}
is the tight-binding dispersion, $\alpha$ is the Rashba spin-orbit interaction, $\Delta_0$ is the $s$-wave superconducting order parameter, and  $J$ is the magnetic exchange coupling between the ordered moments of magnitude $S$ and the conduction electrons. As previously discussed \cite{Rachel2017,Mascot2022}, we assume the spins to be static in nature due to the hard superconducting gap which suppresses Kondo screening \cite{Balatsky2006,Heinrich2018}.  \\

To introduce a superconducting pairing term with proper Debye energy cut-off, we first diagonalize the Hamiltonian in the normal state. Making use of the unitary transformation
\begin{align}
    d_{\vb{k},j}=\sum_{\sigma}u_{\vb{k},j\sigma} c_{\vb{k},\sigma}
    \label{eq:unitary}
\end{align}
we obtain from Eq.(\ref{eq:HSC})
\begin{align}
    H_{SC}&=\sum_{\vb{k},j=1,2}E^{N}_{\vb{k},j}d^{\dagger}_{\vb{k},j}d_{\vb{k},j}
-\sum_{\vb{k}}\sum_{j,l=1,2}\Delta_{\vb{k},jl}\qty(d^{\dagger}_{\vb{k},j}d^{\dagger}_{-\vb{k},l}+d_{\vb{k},l}d_{-\vb{k},j})
\label{eq:Hband}
\end{align}
where $d^{\dagger}_{\vb{k},j}$ creates an electron with momentum $\vb{k}$ in band $j=1,2$ in the normal state, $E^{N}_{\vb{k},j}$ is the normal state dispersion of band $j$, and $\Delta_{\vb{k},jl}=\Delta_{0}u_{\vb{k},j\uparrow}u_{-\vb{k},l\downarrow}$ is the intra-band (inter-band) superconducting order parameter for $j=l$ $(j \not = l)$.\\

To introduce a Debye energy cut-off, we assume that only electrons in an energy range of $\omega_D$ around the the Fermi energy, i.e., for $|E^{N}_{\vb{k},j}|<\omega_D$, participate in the pairing process. Thus we introduce the Debye energy as follows
\begin{align}
\Delta_{\vb{k},jl}\qty(\omega_{D})&=\Delta_{\vb{k},jl}\theta\qty(\omega_{D}-\abs{E_{\vb{k},j}^{N}})\theta\qty(\omega_{D}-\abs{E_{\vb{k},l}^{N}})
\end{align}

\section{Derivation of $P^{(2)}_{SC}$ for a $d_{x^2-y^2}$-wave superconductor}
As derived in Ref.\cite{Mahmood2022}, the two electron photoemission counting rate arising from the emission of a Cooper pair whose electrons possess opposite momenta and spins, is given by
\begin{equation}
    P^{(2)}_{SC}\qty(\omega_{\vb{q}},\vb{k}_1')=2\pi\delta\qty(\omega_{\vb{q}}-2\epsilon_{\vb{k}_1'})\abs{ \sum_{\vb{k}}\frac{\gamma_0 V\qty(\vb{k}-\vb{k}_1')}{\omega_{\vb{q}}-E_{\vb{k}}-\epsilon_{\vb{k}}}\frac{\Delta_{\vb{k}}}{2E_{\vb{k}}}}^2
    \label{eq:P2SC}
\end{equation}
where $E_{\vb{k}}=\sqrt{\xi_{\vb{k}}^2+\Delta_{\vb{k}}^2}$ is the energy dispersion in the superconducting state and $\epsilon_{\vb{k}}=\frac{\hbar^2k^2}{2m_e}+W$ is sum of the kinetic energy of the photoelectron with momentum $\vb{k}$ and the work function $W$. \\

Since the Coulomb scattering occurs between a photoelectron and a conduction electron, we can assume that it is weakly screened and we therefore confine the integration to a small region of radius $\kappa$ ($\kappa^{-1}$ being the screening length) around $\vb{k}_1^\prime$ in which we assume that the integrand varies only weakly.  We then obtain
\begin{equation}
    P^{(2)}_{SC}\qty(\vb{k}_1') \approx 2\pi\delta\qty(\omega_{\vb{q}}-2\epsilon_{\vb{k}_1'}) \abs{ \gamma_0 \pi \kappa^2 \frac{V_0}{\kappa^2} \frac{1}{\omega_{\vb{q}}-E_{\vb{k}_1^\prime}-\epsilon_{\vb{k}_1^\prime}}\frac{\Delta_{\vb{k}_1^\prime}}{2E_{\vb{k}_1^\prime}}}^2
    \label{eq:P2SC1}
\end{equation}

Since $\omega_{\vb{q}}=2\epsilon_{\vb{k}_1'}$ in Eq.(\ref{eq:P2SC1}), we also have
\begin{align}
    \omega_{\vb{q}}-E_{\vb{k}_1^\prime}-\epsilon_{\vb{k}_1^\prime}= \frac{\omega_{\vb{q}}}{2}-E_{\vb{k}} \approx \frac{\omega_{\vb{q}}}{2}
\end{align}
where the last approximation is valid since the relevant energies of the conduction electrons are on the order of 10-100 meV, while those of the photons are of the order of tens of eV. Thus we can write Eq.(\ref{eq:P2SC1}) as
\begin{align}
    P^{(2)}_{SC}\qty(\vb{k}_1') \approx & 2\pi\delta\qty(\omega_{\vb{q}}-2\epsilon_{\vb{k}_1'}) \frac{\pi^2\gamma_0^2 V_0^2}{\omega^2_{\bf q}}
    \frac{\Delta^2_{\vb{k}_1^\prime}}{\xi^2_{{\bf k}_1^\prime}+\Delta^2_{\vb{k}_1^\prime}} \nonumber \\
   =& P^{(2)}_{SC}\qty(\vb{k}_F)\frac{\Delta^2_{\vb{k}_1'}}{\xi_{\vb{k}_1'}^2+\Delta^2_{\vb{k}_1'}}
    \label{eq:P2SC2}
\end{align}
which is Eq.(4) of the main text.

\section{Emergence of superconducting correlations for inter-band pairing only in the 2DTSC model}

Introducing the Nambu spinor
\begin{align}
    \psi^{\dagger}_{\bf k}=\begin{pmatrix}
    d^{\dagger}_{\vb{k},1}, &  d^{\dagger}_{\vb{k},2}, &  d_{-\vb{k},2}, & d_{-\vb{k},1}
\end{pmatrix}
\end{align}
we can write the Hamiltonian in Eq.(\ref{eq:Hband}) as
\begin{align}
    H_{SC} = \sum_{\bf k} \psi^{\dagger}_{\bf k} {\hat H}_{BdG} \psi_{\bf k}
\end{align}
where ${\hat H}_{BdG}$ is the Bogoliubov-de Gennes (BdG) Hamiltonian matrix in the band basis given by
\begin{align}
    {\hat H}_{BdG}&=\begin{pmatrix}
        E^{N}_{\vb{k},1} & 0 & -\Delta_{\vb{k},12} & -\Delta_{\vb{k},11}\\
        0 & E^{N}_{\vb{k},2} & -\Delta_{\vb{k},22} & -\Delta_{\vb{k},21}\\
        -\Delta_{\vb{k},12}^{*} & -\Delta_{\vb{k},22}^{*} & -E^{N}_{-\vb{k},2} & 0\\
        -\Delta_{\vb{k},11}^{*} & -\Delta_{\vb{k},21}^{*} & 0 & -E^{N}_{-\vb{k},1}
    \end{pmatrix}
    \label{eq:Hmatrix}
\end{align}

Considering first the case of intra-band pairing only (i.e., no inter-band pairing), we set $\Delta_{\vb{k},jl}=0$ if $j \not = l$. The Hamiltonian matrix then becomes
\begin{align}
    {\hat H}_{BdG}&=\begin{pmatrix}
        E^{N}_{\vb{k},1} & 0 & 0 & -\Delta_{\vb{k},11}\\
        0 & E^{N}_{\vb{k},2} & -\Delta_{\vb{k},22} & 0\\
        0 & -\Delta_{\vb{k},22}^{*} & -E^{N}_{-\vb{k},2} & 0\\
        -\Delta_{\vb{k},11}^{*} & 0 & 0 & -E^{N}_{-\vb{k},1}
    \end{pmatrix}
\end{align}
Using the spinor
\begin{align}
    \psi^{\dagger}_{\bf k}=\begin{pmatrix}
    d^{\dagger}_{\vb{k},1}, & d_{-\vb{k},1}, & d^{\dagger}_{\vb{k},2}, & d_{-\vb{k},2}
\end{pmatrix}
\end{align}
we can rewrite the Hamiltonian in a block-diagonal form as follows
\begin{align}
    {\hat H}_{BdG}&=\begin{pmatrix}
        E^{N}_{\vb{k},1} & -\Delta_{\vb{k},11} & 0 & 0\\
        -\Delta_{\vb{k},11}^{*}  & -E^{N}_{-\vb{k},1} & 0 & 0 \\
        0 & 0 & E^{N}_{\vb{k},2} & -\Delta_{\vb{k},22} \\
        0 & 0 & -\Delta_{\vb{k},22}^{*} & -E^{N}_{-\vb{k},2}
    \end{pmatrix}
\end{align}
This Hamiltonian matrix represents two independent single band superconductors, and its blocks thus possess the same structure as the Hamiltonian matrix for the $d_{x^2-y^2}$-wave superconductor. \\

Conversely, the case of inter-band pairing only can be considered by setting $\Delta_{\vb{k},11}=\Delta_{\vb{k},22}=0$, in which case we obtain from the Hamiltonian matrix in Eq.(\ref{eq:Hmatrix})
\begin{align}
    H_{BdG}&=\begin{pmatrix}
        E^{N}_{\vb{k},1} & 0 & -\Delta_{\vb{k},12} & 0\\
        0 & E^{N}_{\vb{k},2} & 0 & -\Delta_{\vb{k},21}\\
        -\Delta_{\vb{k},12}^{*} & 0 & -E^{N}_{-\vb{k},2} & 0\\
        0 & -\Delta_{\vb{k},21}^{*} & 0 & -E^{N}_{-\vb{k},1}
    \end{pmatrix}
\end{align}
This Hamiltonian matrix can be written in block-diagonal form using the spinor
\begin{align}
    \psi^{\dagger}_{\bf k}=\begin{pmatrix}
    d^{\dagger}_{\vb{k},1}, & d_{-\vb{k},2}, & d^{\dagger}_{\vb{k},2}, & d_{-\vb{k},1}
\end{pmatrix}
\end{align}
which yields
\begin{align}
    H_{BdG}&=\begin{pmatrix}
        E^{N}_{\vb{k},1} &  -\Delta_{\vb{k},12} & 0 & 0\\
        -\Delta_{\vb{k},12}^{*} & -E^{N}_{-\vb{k},2}  & 0 & 0 \\
        0 & 0 & E^{N}_{\vb{k},2} & -\Delta_{\vb{k},21}\\
        0 & 0& -\Delta_{\vb{k},21}^{*} & -E^{N}_{-\vb{k},1}
    \end{pmatrix}
    \label{eq:block}
\end{align}
Considering the block on the upper left, we obtain
 \begin{align}
    {\hat \beta}^\dagger_{\vb{k}}  \begin{pmatrix}
        E^{N}_{\vb{k},1} & -\Delta_{\vb{k},12}\\
        -\Delta_{\vb{k},12}^{*} & -E^{N}_{-\vb{k},2}
    \end{pmatrix} {\hat \beta}_{\vb{k}} =\begin{pmatrix}
        E_{\vb{k},1} & 0 \\
        0 & -E_{-\vb{k},2}
    \end{pmatrix}
\end{align}
where ${\hat \beta}_{\vb{k}}$ represents the Bogoliubov transformation
\begin{align}
    \begin{pmatrix}
        d_{\vb{k},1}\\
        d^{\dagger}_{-\vb{k},2}
    \end{pmatrix}=\begin{pmatrix}
        \beta_{\vb{k},11} & \beta_{\vb{k},12}\\
        \beta_{\vb{k},21} & \beta_{\vb{k},22}
    \end{pmatrix}\begin{pmatrix}
        \gamma_{\vb{k},1}\\
        \gamma_{-\vb{k},2}^{\dagger}
    \end{pmatrix}\equiv {\hat \beta}_{\vb{k}}\begin{pmatrix}
        \gamma_{\vb{k},1}\\
        \gamma_{-\vb{k},2}^{\dagger}
    \end{pmatrix}
    \label{eq:Bogoliubov}
\end{align}
where $\gamma^{\dagger}_{\vb{k},j}$ creates a quasiparticle state with energy $E_{\vb{k},j}$ and $\beta_{\vb{k},ij}$ are the corresponding coherence factors. Analogous results are obtained when diagonalizing the block on the lower right of Eq.(\ref{eq:block}).\\

Next, we investigate the form of the superconducting correlation $\expval{c^{\dagger}_{\vb{k},\sigma_1}c^{\dagger}_{-\vb{k},\sigma_2}}$ in the case of inter-band pairing only. Using the unitary transformation of Eq.(\ref{eq:unitary}), we obtain
\begin{align}
\expval{c^{\dagger}_{\vb{k},\sigma_1}c^{\dagger}_{-\vb{k},\sigma_2}}&=\sum_{j,l}u_{\vb{k},j\sigma_1}u_{-\vb{k},l\sigma_2} \expval{d^{\dagger}_{\vb{k},j}d^{\dagger}_{-\vb{k},l}} \nonumber\\
   &=u_{\vb{k},1\sigma_1}u_{-\vb{k},2\sigma_2} \expval{d^{\dagger}_{\vb{k},1}d^{\dagger}_{-\vb{k},2}}
   +u_{\vb{k},2\sigma_1}u_{-\vb{k},1\sigma_2} \expval{d^{\dagger}_{\vb{k},2}d^{\dagger}_{-\vb{k},1}} \ .
   \label{eq:SCcorr}
\end{align}
Using next the Bogoliubov tranformation in Eq.(\ref{eq:Bogoliubov}), we obtain
\begin{align}
    C_{12}\qty(\vb{k})&\equiv\expval{d^{\dagger}_{\vb{k},1}d^{\dagger}_{-\vb{k},2}}=\beta_{\vb{k},11}^{*}\beta_{\vb{k},21}\expval{\gamma^{\dagger}_{\vb{k},1}\gamma_{\vb{k},1}}+\beta_{\vb{k},12}^{*}\beta_{\vb{k},22}\expval{\gamma_{-\vb{k},2}\gamma^{\dagger}_{-\vb{k},2}} \nonumber \\
  C_{21}\qty(\vb{k}) &\equiv   \expval{d^{\dagger}_{\vb{k},2}d^{\dagger}_{-\vb{k},1}} =\beta_{-\vb{k},21}\beta_{-\vb{k},11}^{*}\expval{\gamma_{-\vb{k},1}\gamma^{\dagger}_{-\vb{k},1}}+\beta_{-\vb{k},22}\beta_{-\vb{k},12}^{*}\expval{\gamma^{\dagger}_{\vb{k},2}\gamma_{\vb{k},2}}
\end{align}
where
\begin{equation}
\beta^{*}_{\vb{k},11}\beta_{\vb{k},21}=-\beta^{*}_{\vb{k},12}\beta_{\vb{k},22}=\frac{\Delta_{\vb{k},12}^{*}}{2\sqrt{\left(\frac{E^{N}_{\vb{k},1}+E^{N}_{-\vb{k},2}}{2}\right)^2+\abs{\Delta_{\vb{k},12}}^2}}
\label{eq:cohfac}
\end{equation}
For momentum satisfying $E_{\vb{k},1}<0<E_{-\vb{k},2}$, the superconducting correlation functions at zero temperature are then given by
\begin{align}
    \expval{c^{\dagger}_{\vb{k},\sigma_1}c^{\dagger}_{-\vb{k},\sigma_2}}&=u_{\vb{k},1\sigma_1}u_{-\vb{k},2\sigma_2}\qty(\beta_{\vb{k},11}^{*}\beta_{\vb{k},21}+\beta_{\vb{k},12}^{*}\beta_{\vb{k},22})=u_{\vb{k},1\sigma_1}u_{-\vb{k},2\sigma_2}\qty[\beta_{\vb{k}}\beta_{\vb{k}}^{\dagger}]_{21}=0
\end{align}
where the last equation holds due to the unitarity of ${\hat \beta}_{\vb{k}}$. This implies that superconducting correlations are absent, and hence Cooper pairs do not form, for those momenta where one of the bands is occupied ($E_{\vb{k},j}<0$), while the other one is unoccupied ($E_{\vb{k},l}>0$). The momentum dependence of $C_{12}\qty(\vb{k})$ together with that of $\beta_{\vb{k},11}^{*}\beta_{\vb{k},21}$ is shown in Supplementary Fig.~\ref{fig:SIFig1}.
\begin{figure}[htb]
\center
\includegraphics[width=0.6\linewidth]{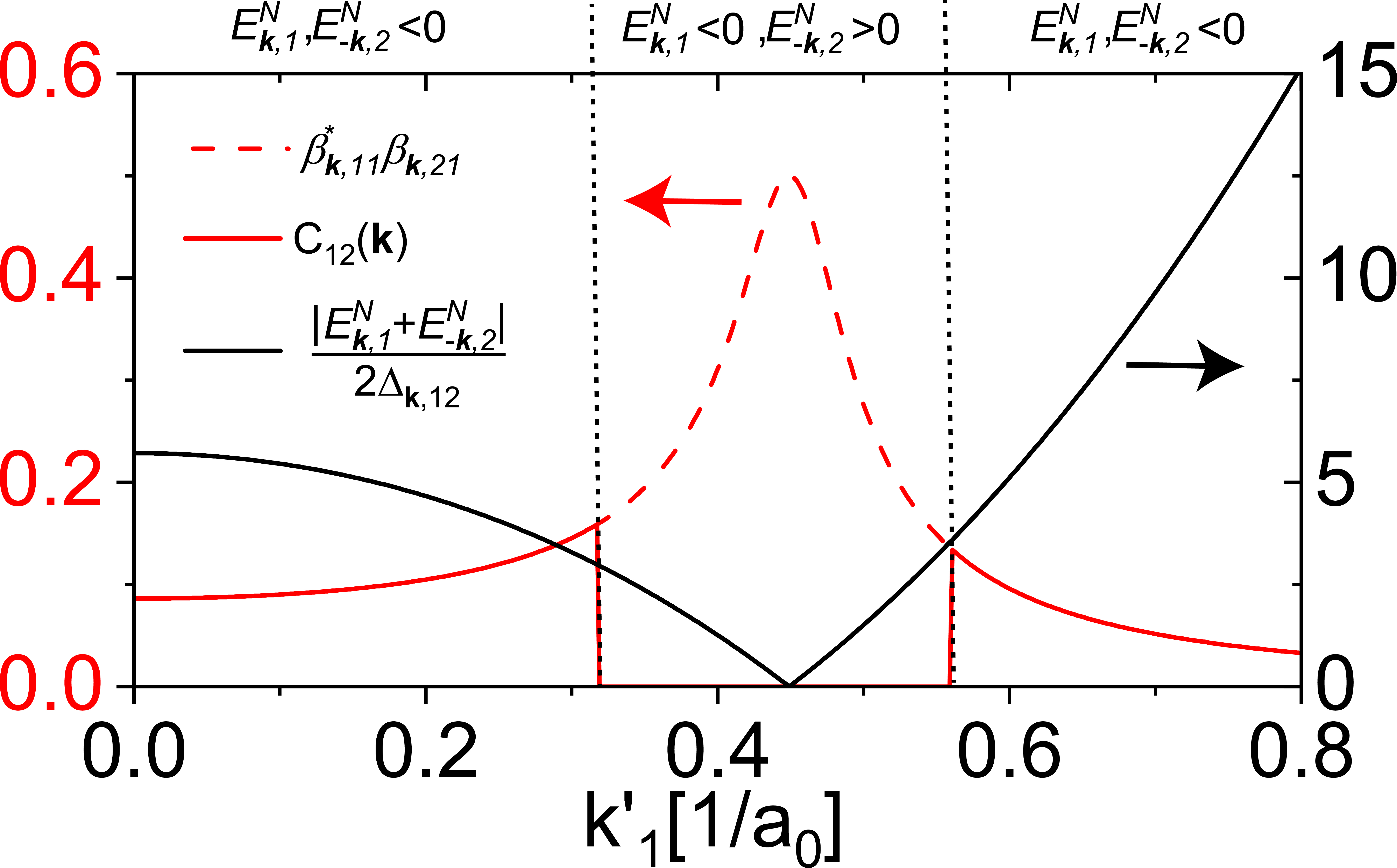}
\caption{Momentum dependence of the superconducting interand correlations $C_{12}(\vb{k})$ (red solid line), and of the product $\beta^{*}_{\vb{k},11}\beta_{\vb{k},21}$ (red dotted line), and of the ratio $(E^{N}_{\vb{k},1}+E^{N}_{-\vb{k},2})/(2\Delta_{\vb{k},12})$ (solid black line).  Parameters are $(t,\mu,\alpha,\Delta,J)=\qty(200,0,10,7,20)\mathrm{meV}.$ }
\label{fig:SIFig1}
\end{figure}
The vertical black dotted lines indicate the momenta, when one of the bands crosses zero energy, and $C_{12}({\bf k})$ vanishes since $\beta_{\vb{k},12}^{*}\beta_{\vb{k},22} = - \beta_{\vb{k},11}^{*}\beta_{\vb{k},21}$. We also find that $\beta_{\vb{k},11}^{*}\beta_{\vb{k},21}$ (see red dashed line) exhibits a maximum when $E^N_{{\bf k},1}+E^N_{-{\bf k},2}$ (see black line) vanishes, as expected from the analytical form of the $\beta_{\vb{k},11}^{*}\beta_{\vb{k},21}$ in Eq.(\ref{eq:cohfac}).

The fact that $P^{(2)}_{SC}$ reflects the form of the superconducting correlations (for details, see Ref.\cite{wong2eARPES}) also explains the qualitatively different form of $P^{(2)}_{SC}$ for inter-band and inter-band pairing, as shown in Fig.3(c) of the main paper. While $P^{(2)}_{SC}$ for intra-band pairing essentially follows the same form as shown in Eq.(4) of the main text, and thus rapidly decays away from the Fermi momentum, $P^{(2)}_{SC}$ for inter-band pairing involves the coherence factors shown in Eq.(\ref{eq:SCcorr})-(\ref{eq:cohfac}), thus decreasing much slower away from the Fermi momentum due to the weaker increase in $E^N_{{\bf k},1}+E^N_{-{\bf k},2}$, as shown in Supplementary Fig.~\ref{fig:SIFig1}.

\section{$P^{(2)}_{SC}$ in the spin triplet channel for the 2DTSC model}
We found in the main text, that the form of $P^{(2)}_{SC}$ in the spin triplet channel is much more symmetric around the Fermi momenta $k^{(1,2)}_{F}$ than in the spin-singlet channel. The reason for this difference lies in the significantly suppressed inter-band pairing in the spin triplet case, as shown in Supplementary Fig.~\ref{fig:SIFig2}.
\begin{figure}[htb]
\center
\includegraphics[width=0.6\linewidth]{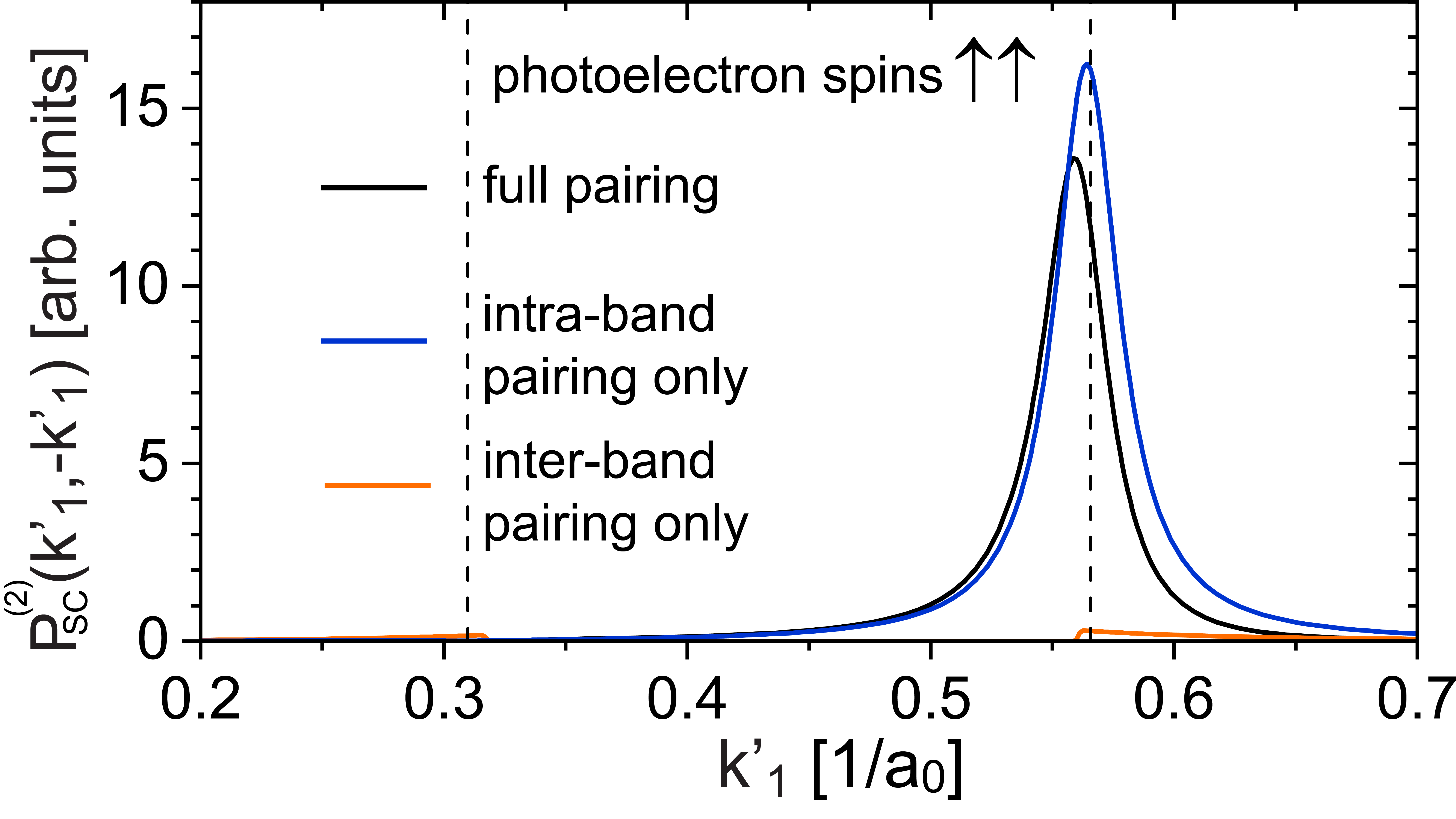}
\caption{$P^{(2)}_{SC}$ in the $\uparrow \uparrow$ triplet channel assuming full (black line), intra-band (blue line) and inter-band (orange line) pairing only. Parameters used are $(t,\mu,\alpha,\Delta,JS)=(200,-760,10,7,20)\mathrm{meV}$.}
\label{fig:SIFig2}
\end{figure}
Here, we present the momentum dependence of $P^{(2)}_{SC}$ computed using the full pairing, intra-band pairing, or inter-band pairing only. The greatly suppressed inter-band pairing in the triplet channel arises from the strong and opposite spin-polarization of the bands, shown in Fig.4(c) of the main text, thus leading to a more symmetric form of $P^{(2)}_{SC}$ in the triplet channel. As previously pointed out \cite{wong2eARPES}, the correlations in the triplet $S_z=0$ channel vanish identically.

\section{$P^{(2)}_{SC}$ in the topological $C=2$ phase of the 2DTSC model}

Finally, we briefly discuss the form of $P^{(2)}_{SC}$ for a set of parameters, when the system in the topological $C=2$ phase \cite{Rachel2017}. As previously discussed, $P^{(2)}_{SC}$ cannot be employed to distinguish between the topological and the trivial phases of the system, though it can detect the topological phase transition \cite{wong2eARPES}.
\begin{figure}[htb]
\center
\includegraphics[width=0.75\linewidth]{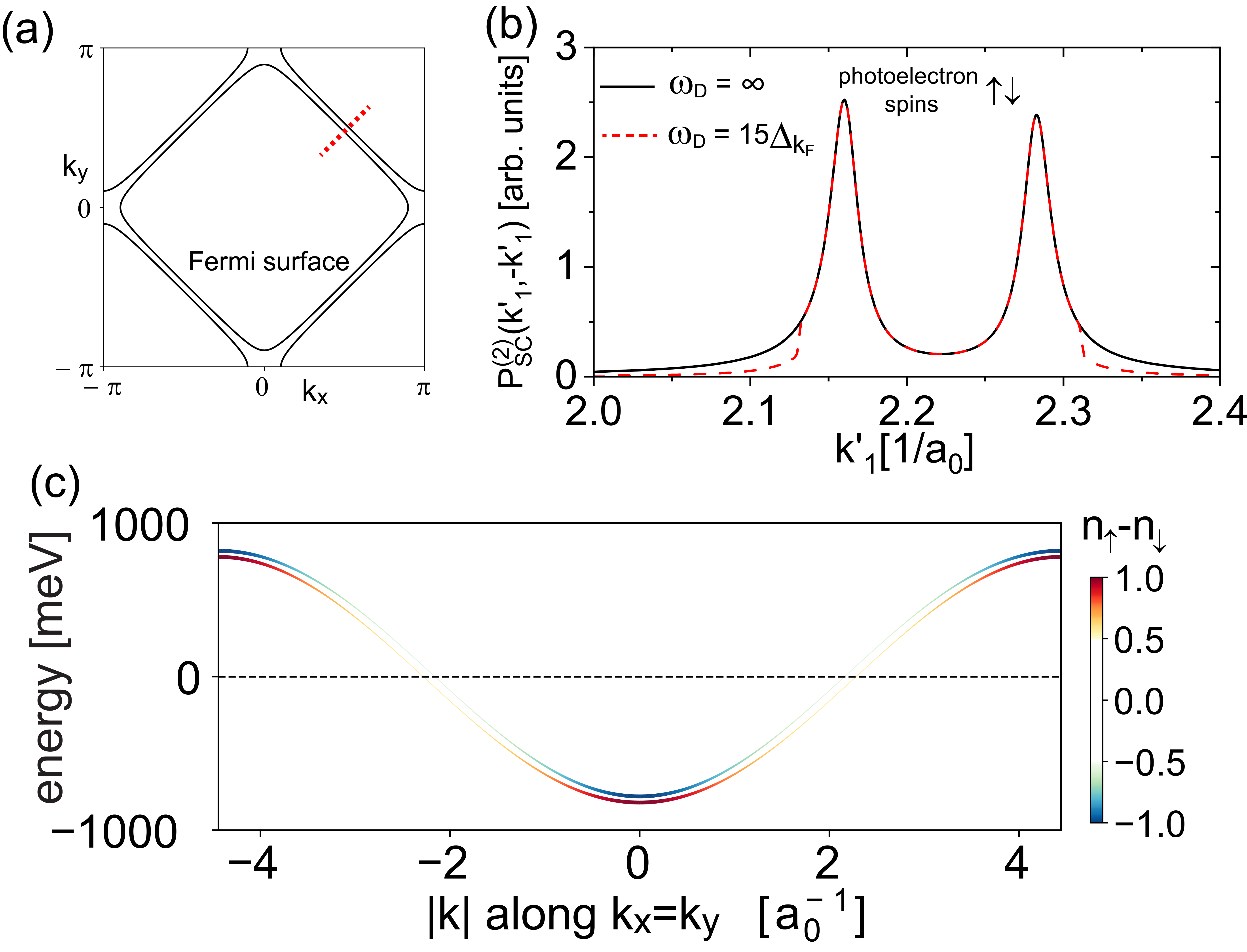}
\caption{(a) Fermi surfaces in the normal state (b) $P^{(2)}_{SC}$ in the spin singlet channel as a function of ${\bf k}_1^\prime$ along the diagonal of the BZ, as indicated by the red dotted line in (a), without Debye cutoff (black line) and with Debye cutoff $\omega_D=15\Delta_{\vb{k}_F}$ (red line). (c) Spin polarization of the electronic bands in the normal state.  Parameters used are $(t,\mu,\alpha,\Delta,JS)=(200,0,10,7,20)\mathrm{meV}$.}
\label{fig:SIFig3}
\end{figure}
For this parameter set, the Fermi surfaces shown in Supplementary Fig.~\ref{fig:SIFig3}(a), and the resulting $P^{(2)}_{SC}$ as a function of momentum along $k_x=k_y$ [see red dotted line in Supplementary Fig.~\ref{fig:SIFig3}(a)] is shown in Supplementary Fig.~\ref{fig:SIFig3}(b). In contrast to the results shown in Fig.3(b) of the main text, the form of $P^{(2)}_{SC}$ is much more symmetric around $k^{(1,2)}_F$, which is due to the much weaker spin-polarization of the bands around the Fermi energy [see Supplementary Fig.~\ref{fig:SIFig3}(c)], and thus a much stronger intra-band than inter-band pairing.